\documentclass[12pt,epsf,amssymb]{article}
\usepackage{tabularx}
\usepackage{array}
\usepackage{graphics}
\usepackage{graphicx}
\usepackage{epsfig}
\usepackage{amsmath}
\usepackage{amssymb}
\usepackage{citesort}
\makeatletter

\usepackage{verbatim}

\newcommand{\bea}{\begin{eqnarray}}
\newcommand{\eea}{\end{eqnarray}}
\newcommand{\beq}{\begin{equation}}
\newcommand{\eeq}{\end{equation}}
\newcommand{\gev}{{\rm GeV}}
\newcommand{\mev}{{\rm MeV}}

\newcommand{\pdir}{p\kern -5.2pt\raise 0.2ex\hbox {/}}
\newcommand{\vdir}{v\kern -5.75pt\raise 0.15ex\hbox {/}}
\newcommand{\kdir}{k\kern -5.75pt\raise 0.15ex\hbox {/}}
\newcommand{\epsdir}{\epsilon\kern -5.0pt\raise 0.15ex\hbox {/}}
\newcommand{\bvdir}{\bar{v}\kern -5.75pt\raise 0.15ex\hbox {/}}
\newcommand{\Ddir}{D\kern -7.75pt\raise 0.20ex\hbox {/}}
\newcommand{\Adir}{A\kern -7.75pt\raise 0.20ex\hbox {/}}
\newcommand{\ldir}{l\kern -5.0pt\raise 0.2ex\hbox{/}}
\newcommand{\varepsdir}{\varepsilon\kern -5.5pt\raise 0.15ex\hbox{/}}

\newcommand{\gbb}{g_{{B^\ast B \pi}}}
\newcommand{\gdd}{g_{{D^\ast D \pi}}}
\newcommand{\ghh}{g_{{H^\ast H \pi}}}
\newcommand{\ghat}{\widehat g}
\newcommand{\mbs}{m_{B^\ast}}

\makeatother

\begin{document}
\thispagestyle{empty}

\begin{flushright}
\begin{tabular}{l}
{\tt LPT Orsay, 03-72}\\
{\tt SWAT/386}
\end{tabular}
\end{flushright}
\vskip 2.6cm\par
\begin{center}
{\par\centering \textbf{\LARGE Lattice measurement } }\\
\vskip .45cm\par
{\par\centering \textbf{\LARGE of the couplings $\ghat_\infty$ and $\gbb$ } }\\
\vskip 0.9cm\par
{\par\centering 
\sc A.~Abada$^a$, D.~Be\'cirevi\'c$^a$, Ph.~Boucaud$^a$, G.~Herdoiza$^{a,b}$,
J.P.~Leroy$^a$, A.~Le~Yaouanc$^a$, O.~P\`ene$^a$}
{\par\centering \vskip 0.5 cm\par}
{\par\centering \textsl{
Laboratoire de Physique Th\'eorique (B\^at.210), Universit\'e de
Paris XI,\\
Centre d'Orsay, 91405 Orsay-Cedex, France.} \\
\vskip 0.3cm\par}
{\par\centering \textsl{
Department of Physics, University of Wales Swansea,\\
Singleton Park, Swansea, SA2 8PP, United Kingdom.} \\
\vskip 0.6cm\par}
\today
\end{center}

\vskip 0.45cm \begin{abstract}
We present the results of a quenched lattice QCD study of the coupling $\ghat$, 
in the static heavy quark limit. After combining this with our previous 
results obtained by using propagating heavy quarks with a mass around 
the physical charm quark, we are able to interpolate to the $b$-quark sector. 
Our results are $\ghat_\infty = 0.48 \pm 0.03\pm 0.11$, $\ghat_b = 0.58 \pm 0.06 \pm
0.10$, and $\gbb = 47 \pm 5 \pm 8$. 
\end{abstract}
\vskip 2.2 cm

\setcounter{page}{1}
\setcounter{equation}{0}

\renewcommand{\thefootnote}{\arabic{footnote}}
\vspace*{-1.5cm}
\newpage
\setcounter{footnote}{0}
\section*{Introduction}

\hspace*{\parindent}The coupling of the vector and 
pseudoscalar heavy-light mesons to the pion, $\ghh$ , and the related $\ghat_Q$ defined through\footnote{The subscript Q $\in\,\{c,b\}$ refers to the heavy quark constituent of the heavy-light meson.}
\bea \label{gg}
\ghh \ = \  \frac{2 \sqrt{m_H m_{H^\ast}}}{f_\pi} \ \ghat_Q\,,
\eea
have been extensively studied in the literature, both for phenomenological
and theoretical reasons (a list of results can be found in refs.~\cite{QCDSR,AD,Singer,gdd,Colangelo_0201305}).
$\ghat_Q$ is particularly important in the studies of the $q^2$-dependence of
the form factor $F_+(q^2)$ which encodes the non-perturbative QCD dynamics of $D\to \pi$ and $B\to \pi$
semileptonic decays. To better illustrate this, we consider the dispersion formula for  $F_+^{B\to \pi}(q^2)$ : 
\bea  \label{Fp}
F_+^{B\to \pi}(q^2)\ = \ { \frac{1}{2\mbs}}\  \frac{f_{B^*} \gbb}{1 \ -\ \displaystyle{\frac{q^2}{m_{B^*}^2}}}  + 
\frac{1}{\pi}
\int_{(m_\pi + m_B)^2}^{\infty} \!\! dt\; \frac{{\rm Im} F_+(t)}{t-q^2-i\varepsilon}   \; ,
\eea
where $m_{B^*}$ and $f_{B^*}$ are the mass and the decay constant of the $B^\ast$-meson. 
Near  the  $B^\ast$-pole,  $F_+$ can be approximated by the
first  term of the r.h.s of eq.~(\ref{Fp}), which is why $\gbb$ is so important. Inversely, from 
the $q^2$-dependence deduced from the lattice data for $F_+^{B\to \pi}$, one can extract indirectly 
the value of $\gbb$~(see e.g.~\cite{Abada:2000ty}).
Moreover the need for a more accurate estimate of $\ghat_Q$  became recently more evident since  
the guidance of the chiral extrapolations by the predictions of the heavy meson chiral 
perturbation theory necessarily involves $\hat g_Q$~\cite{chiralB}. In the effective theory 
based on the combination of the heavy quark and
chiral symmetries~\cite{HQC,casalb} one of the  essential parameters is the coupling $\ghat_\infty$
which appears as the infinite quark mass limit of $\ghat_Q$ :
\bea
\label{gm}
\ghat_Q = \ghat_{\infty} + \mathcal{O}\left( {1/m_Q^n} \right)\,.
\eea

While the $B^\ast \to B\pi$ decay is forbidden by the lack of phase space and therefore the coupling 
$\gbb$ cannot be measured in experiment, the coupling $\gdd$ has already been measured. CLEO reported recently~\cite{CLEO}:
\bea \label{CLEO}
\gdd &=& 17.9\pm 0.3\pm 1.9\;,\\
{\it i.e} \quad \ghat_c^{\ \rm exp.} &=& 0.61 \pm 0.01 \pm 0.07\;, \eea
where we used eq.~(\ref{gg}) to obtain $\ghat_c$ from $\gdd$. 
Since the pion emerging from this decay is soft, the experimental measurement of $\Gamma(D^{\ast +})$  
is difficult. The experimental confirmation of the CLEO results would be highly welcome.

Regarding the theoretical estimates the situation is still quite unsatisfactory.
Typical values for $\ghat_Q$, as obtained by using the QCD sum rules (QCDSR) are
low, around $0.3$~\cite{QCDSR}, in disagreement with $\ghat_c^{\ \rm exp.}$. Such a 
disagreement with the experimental value was not, however, observed
with similar couplings, such as $g_{\rho\omega\pi}$, $g_{NN\pi}$, 
or $g_{\Sigma_c^\ast\Lambda_c\pi}$. To overcome that problem, in ref.~\cite{Becirevic:2002vp}
it has been proposed to include the first radial excitations in the hadronic part of the 
sum rules, below the duality threshold. As a result 
the coupling $\ghat_c$ becomes much larger ($\ghat_c \sim 0.7$), while the similar 
modification of the sum rule leaves the decay constants  ($f_D$ and $f_{D^\ast}$) unchanged.

Predictions of various {quark models} are in the range 
$ 0.3 \lesssim \ghat_{\infty} \lesssim  0.8$~\cite{QM}. In particular, in the model with the 
Dirac equation a value of $\ghat_\infty$ around 0.6 was
found before the experimental result became available~\cite{AD}.
On the {lattice}, a quenched calculation by UKQCD~\cite{UKQCD} led to $\ghat_\infty =
0.42(4)(9)$. This exploratory study was carried out in the static heavy quark
limit where the difficulties to isolate the ground state are well known.
In addition it was performed on a coarse lattice, with a low 
statistics and with only two light quark masses from which it was extrapolated to
the chiral limit. Finally, the indirect measurements using the lattice data for the form
factors $F_+$ and $F_0$ \cite{Abada:2000ty} have extracted rather small values~\cite{Abada:2000ty}.

In our recent paper~\cite{gdd}, we presented the results of the quenched 
lattice simulation in which we used the propagating heavy quarks with the mass 
around the physical charm quark. We obtained, $\ghat_c = 0.67 \pm 0.08^{+0.04}_{-0.06}$, 
and observed no apparent dependence on the heavy quark mass. 
Actually the  3 values of  $\ghat_Q$ measured directly in the neighbourhood of the $c$-quark mass
were compatible within our error bars and shew only a very small negative slope (see section 4.3 of ref.~\cite{gdd}). This situation allowed
a safe {\it interpolation} to the  $c$-quark but  the determination of $\ghat_\infty$ relied heavily  on the hypothesis that
the observed behaviour was valid over the whole mass-range. In order to  check the validity of this asumption we have performed  a lattice quenched computation of $\ghat$ In the static heavy
quark limit, the outcome of which is  presented  
in this paper.    Our result is
\bea \label{ginf}
\ghat_\infty = 0.48 \pm 0.03\pm 0.11 \;,
\eea
where the errors are statistical and systematic, respectively. 
Using this and the results we presented in ref.~\cite{gdd}, we were able to 
interpolate to the $b$-quark. We get
\bea \label{gb}
\ghat_b &=& 0.58 \pm 0.06 \pm 0.10 \,, \\
\gbb &=& 47 \pm 5 \pm 8 \,. \nonumber
\eea
By comparing eqs.~(\ref{ginf}) and (\ref{gb}), our results suggest that the $1/m_B$
corrections to $\ghat_\infty$ are of the order of $20-30$\%.

The  paper is organized as follows: after this introduction we 
explain in Sec.~\ref{sec2} the relation between $\ghat$ and the matrix elements that we
compute on the lattice; in Secs.~\ref{sec3} and~\ref{sec3bis} we present the details of our 
lattice study and show our main results; Sec.~\ref{syst} is devoted to 
the discussion of systematic errors; in Sec.~\ref{interp} we   interpolate to the physical $B$-meson mass; finally we summarize our findings in Sec.~\ref{sec4}.

\section{Extracting $\ghat_\infty$\label{sec2}}

\hspace*{\parindent}The coupling $\ghh$ (with $H\in \{D,\ B\}$) is defined by the matrix element 
\bea
\langle\,
H(p^{\prime})\,\pi(q)\,\vert \, H^\ast(p,\,\lambda)\,\rangle = \ghh \ \left( q\cdot \epsilon^\lambda(p)\right)\; ,
\eea 
where $q = p - p^{\prime}$ and  $\epsilon^\lambda(p)$ is the polarization of the vector meson. To determine $\ghh$, we can consider the matrix element 
of the divergence of the light--light axial 
current, $A_\mu = {\overline{q}} \gamma_\mu \gamma_5 q$. In the limit where the pion 
is soft, we can write
\bea \label{dA}
    \langle H(p^{\prime}) | q_{\mu}A^{\mu} | H^{\ast}(p,\,\lambda) \rangle =
    \ghh \frac{q\cdot \epsilon^\lambda(p)}{m_{\pi}^2 - q^2}
    \times
    f_\pi m_{\pi}^2\ +\, \dots
\eea
where the dots
stand for terms suppressed  when $q^2$ is small and close to the $m_\pi^2$-pole.
To bring out $\ghh$ from (\ref{dA}) we express the l.h.s in terms of the form
factors, $A_{0,1,2}\,$, evaluated on the lattice. The standard parametrization
reads
\bea \label{3A}
\left\langle H(p^{\prime})
\left|A^\mu\right|H^{\ast}(p,\lambda)\right\rangle &=&
2m_{H^{\ast}}A_0(q^2)\frac{\epsilon^{\lambda}\cdot q}{q^2}q^\mu  +
(m_{H^{\ast}} +
m_H)A_1(q^2)\left[\epsilon^{\lambda\,\mu}-\frac{\epsilon^\lambda\cdot
q}{q^2}q^\mu\right]
\nonumber \\
&& +  A_2(q^2)\:\frac{\epsilon^\lambda\cdot q}{m_H+m_{H^{\ast}}}\left[p^{\mu}+p^{\prime\,\mu}
-\frac{m_{H^{\ast}}^2-m_H^2}{q^2}q^{\mu}\right] \; .\eea
Note that when ${\vec q} = \vec 0$, the soft pion limit is verified, since $q^2 =
(m_{H^{\ast}}-m_{H})^2 \sim 0$. Moreover, if we choose ${\vec q} = \vec 0$ 
and 
$\vec p = \vec p^{\,\prime} = \vec 0$, eq.~(\ref{3A}) simplifies  to 
\bea \label{A1}
\left\langle H \left|A_i\right|H^{\ast}\right\rangle = (m_{H^{\ast}}+m_{H})\, A_1(0)\,
\epsilon^{\lambda}_{i} \; .
\eea

As discussed in our previous paper~\cite{gdd}, when $q^2 = 0$ the form factors $A_{0,1,2}$ are related 
to one another because the $q^2$-poles in eq.~(\ref{3A}) are unphysical and therefore their residues have to
cancel. This leads to the relation
\bea \label{A012}
2m_{H^{\ast}} A_0(0) = (m_{H^{\ast}}+m_{H}) A_1(0) + (m_{H^{\ast}}-m_{H}) A_2(0) \,.
\eea
When taking the divergence of the axial current in (\ref{3A}), only the term in
$A_0$ remains on the r.h.s. so that by using eq.~(\ref{dA}), we arrive at
\bea \label{gA}
\ghh &=& \frac{2m_{H^{\ast}}A_0(0)}{f_\pi} \\
&=& \frac{1}{f_\pi}\left[ (m_{H^{\ast}}+m_H) A_1(0) + (m_{H^{\ast}}-m_H) A_2(0) \right] \,.
\eea
Finally, inserting the definition~(\ref{gg}) leads to the expression for $\ghat_Q$
\bea \label{gAb}
    \ghat_Q = \frac{(m_{H^{\ast}}+m_H)}{2\sqrt{m_H m_{H^{\ast}}}} A_1(0)+
    \frac{(m_{H^{\ast}}-m_H)}{2\sqrt{m_H m_{H^{\ast}}}} A_2(0)\,.
\eea
Owing to the heavy quark symmetry, the vector and the pseudoscalar heavy--light mesons 
are degenerate in the static limit and eq.~(\ref{gAb}) simply becomes 
\bea \label{gAinf}
    \ghat_\infty = A_1(0) \, .
\eea
Thus, the value of $\ghat_\infty$ (in the static heavy quark limit) is simply given by the form
factor $A_1$ at $\vec q = \vec 0$.

\section{Strategy for the calculation of $\ghat_{_\infty}$ on the lattice\label{sec3}}

\hspace*{\parindent}To determine the form factor $A_1(0)$ from the lattice evaluation of the
matrix element $\left\langle H \left|A^\mu\right|H^\ast \right\rangle$, we use eq.~(\ref{A1}) 
where $H$ and $H^\ast$ now refer to the pseudoscalar and vector mesons containing an
infinitely heavy quark.
We consider the two- and three-point correlation functions, $C_2$ and $C_{3 \,\mu
\nu}$, with local (L) and smeared (S) interpolating fields for 
mesons consisting of a static heavy and a propagating light quarks.

In the case in which both
the source and the sink are smeared, the {two-point} functions are defined as
\bea
\label{C2} C_2^{SS}(t_x) = \left. \langle
\sum_{\vec x}  \ P^S(x) {P^S}^\dagger(0) \rangle \right. \ = \ \frac{1}{3} \ \sum_{i=1}^{3} \ \left.
\langle  \sum_{\vec x}  \ V^S_i(x) {V^S_i}^\dagger (0) \rangle \right.,
\eea
where $P^S$ and $V^S_\mu$ are the smeared interpolating fields of the pseudoscalar
and vector mesons.  
$C_2^{SL}(t_x)$ is obtained by simply replacing $P^S\to P^L=  \bar q \gamma_5 h$,  
$V^S\to V^L=  \bar q \gamma_\mu h$, where $h$ stands for the static heavy quark field.
The equality between the pseudoscalar and vector meson correlation functions in 
eq.~(\ref{C2}) is a consequence of the heavy quark symmetry.
When the excited states are decoupled and the ground state isolated (at large enough $t_x$), 
the two-point function becomes
\bea
\label{CalZ1} C_2^{SS}(t_x) =  ({\cal Z}^S)^2 \, e^{-E_H^{SS}\; t_x} \, ,
\eea
where $E_H$ is the binding energy of the static heavy--light meson, and 
\bea \label{CalZ} {\cal Z}^S \equiv\frac{\left. \langle \, 0 \, \vert P^S \vert\, H \, \rangle \right.}{(2
m_{H})^{1/2}} \; .
\eea

The {\it three-point} Green functions with smeared interpolating fields are defined in the following way,
\bea \label{C3} C_{3 \, \mu \nu}^{SS}(0 ,t_x ,  t_y ) = 
\left. \langle  \sum_{\vec x,\vec y}  \ P^S(y)  A_\nu (x) {V^S_\mu}^\dagger(0)\rangle \right|_{\; 0 < t_x <
t_y}\!\!\!\!\!\!, \eea 
where the vector and pseudoscalar mesons are inserted at the origin and at a fixed
time $t_y$, respectively. The insertion time $t_x$ of the (local) axial current ($A_\mu=\bar q\gamma_\mu \gamma_5 q$) varies in the range
$0\ -\ t_y$. The  matrix element we are considering, 
$\left\langle H \left|A_\mu\right|H^\ast\right\rangle$, is then extracted from the following ratio: 
\bea
\label{ratio} R^{SS}(t_x) &=& \ \frac{1}{3} \ \sum_{i=1}^{3} \  \frac{ C_{3 \,
ii}^{SS}(t_x) {\cal Z}_S {\cal Z}_S}{C_2^{SS}(t_x) C_2^{SS}(t_y-t_x)}  \\
&\rightarrow& {\frac{1}{ 3}} \sum_{i=1}^{3} \epsilon^{\lambda}_{i}(\vec 0) \ \left.\frac{\left\langle H \left|A_i(t_x)\right|H^\ast\right\rangle}{2 m_{H}} \right|_{\; 0 \ll
t_x \ll t_y} \!\!\!\!
\ = \ A_1(0) \ = \ \ghat_\infty\;.\nonumber
\eea
The time dependence cancels
in the ratio, so that a plateau in time of $R^{SS}(t_x)$ directly leads to the value of $A_1(0)$, which for 
$m_q\to 0$ becomes the desired coupling $\ghat_\infty$. In what follows we will use the notation in which 
$A_1(0) \equiv \ghat_\infty$, even though this is strictly true only for $m_q\to 0$.

In our simulation we choose to work with the usual (improved) Wilson light quarks 
and with the static heavy quark by using the Eichten--Hill action~\cite{eichten-hill}. 
In terms of  quark propagators and the Wilson line, the three-point correlation function~(\ref{C3}) 
can be written as 
\bea \label{Wil} C_{3 \, \mu
\nu}^{LL}(t_x) = \left. \langle \; \sum_{\vec x,\vec y}   {\rm Tr} \left[ \frac{1+\gamma_0}{2} \, P_{y}^{0} \, 
\gamma_\mu S_u(0;x;U) \gamma_\nu \gamma_5  S_d(x;y;U) \gamma_5 \right] \, \rangle_{U}
\right.\! , \eea 
where $S_q(x;y;U) =  q(x) \bar q(y)$ is the propagator of the light
quark $q$. In our study the two light quarks are degenerate, $m_u=m_d\equiv m_q$. 
The brackets $\langle \dots \rangle_U$ indicate the average over the background gauge field 
configurations ($U$), whereas $P_{y}^{x}$ is the
ordered product of links in the time direction (Wilson line), 
\bea \label{PU} P_{y}^{\,x}  = \delta({\vec x} -
{\vec y}) \prod\limits_{t_z=t_x}^{t_y-1} U_t(x+t_z\,\hat{t}) \;, \eea
with $\hat t$ being the unit vector in the time direction.

To improve the statistical quality of the signal we implement the recent proposal of
ref.~\cite{DellaMorte:2003mn} and replace the simple link variables in the Wilson line 
by the so called {\it ``fat links"}, defined as
\beq \label{Wilstat} U_t(x)
\rightarrow  U_t^{\rm fat}(x) = \frac{1}{6} \sum\limits_{i=x,y,z} \left[ U^{\rm
Staple}_i( x,x+\hat{t})+  U^{\rm Staple}_{-i}( x,x+\hat{t})
\right] \; . \eeq
In order to isolate the ground state from the correlators~(\ref{Wil}) at the smallest time separations, 
it is necessary to devise an efficient smearing procedure. We use the proposal of ref.~\cite{Boyle:1999gx} 
and replace the quark fields $q(x)$ by  
\beq \label{Smearing}
q(x) \rightarrow \sum\limits_{r=0}^{R_{max}} {\scriptstyle
(r+\frac{1}{2})}^2 \phi({\scriptstyle r})
 \sum\limits_{i=x,y,z} \left\{
\left[ \prod\limits_{k=1}^r U^F_i({\scriptstyle
x+(k-1)\hat{i}})\right] q({\scriptstyle x+r\hat{i}}) + \left[
\prod\limits_{k=1}^r U_i^{F^\dagger}({\scriptstyle
x-k\hat{i}})\right] q({\scriptstyle x-r\hat{i}}) \right\} \;
. \eeq 
The wave function $\phi({r})$ is chosen in such a way that the overlap with the ground state is increased. We take 
$\phi({ r}) = e^{-r/R_b}$, where $R_b$ is a parameter which is fixed by requiring that the smearing be optimal. 
Note that it is not necessary to normalize the wave function since the normalisation factors cancel
 in the ratio~(\ref{ratio}). The smearing also includes the so called fuzzing: 
the gauge links $U^F$ are fuzzed by $N_{F}$ iterations of the following procedure
\beq \label{Fuzzing} U_j(x) \rightarrow U^F_j(x) ={\cal P} \left[
C~U_j(x) +  \sum\limits_{i\neq j}\left[ U^{\rm Staple}_i(
x,x+\hat{j})+  U^{\rm Staple}_{-i}( x,x+\hat{j}) \right] \right]
\; , \eeq 
where ${\cal P}$ is used to project the fuzzed fields onto $SU(3)$, and $C$ is a real parameter.

\section{Conditions and results of the simulation\label{sec3bis}}

\hspace*{\parindent}The main results presented in this work are obtained from the  simulation 
performed on a $24^3 \times 28$ lattice with periodic boundary conditions, 
at $\beta = 6.2$. We collect $160$ independent $SU(3)$ gauge configurations 
in the quenched approximation. The Wilson fermion action is non-perturbatively
$\mathcal{O}(a)$ improved with $c_{SW}= 1.614$~\cite{cswalpha}. The light quark
propagators are computed with the following Wilson hopping
parameters: \bea \label{eq0} \kappa_q &=& 0.1344_{q_1}\ ,\;
0.1348_{q_2}\ ,\; 0.1351_{q_3}\;, \eea 
corresponding to  quark masses around the physical strange quark.

We will present our results using our preferred set of parameters, $t_y = 13\,  a$,
$R_{max} = 5\,a$, $R_b = 3.0\,a$, $N_{F} = 5$ and $C = 4$. The motivation for this choice of
parameters, the effect of the smearing~(\ref{Smearing}) and of the modification 
of the static heavy quark action~(\ref{Wilstat}), will be discussed in 
section~\ref{syst}, where we will also quantify the systematic effects 
involved in our calculation.

\subsection{Study of the two-point functions}

\hspace*{\parindent}{\it Light--light operator:} The axial current appearing in eq.~(\ref{C3})  consists of two 
light quarks. Even though we use the improved Wilson quarks,
for our purpose there is no need to improve the bare axial current since the form factor 
$A_1$ is insensitive to the presence of the ${\cal O}(a)$ improvement term. 
On the other hand,  the improvement 
of the renormalization constant is of course necessary. We use
\bea \label{ZA}
Z_A^I(g_0^2) = Z_A^{(0)}(g_0^2) \left( 1 + \tilde b_A(g_0^2) a \rho \right)\;,
\eea
where the quark mass $a\rho$ is obtained from the axial Ward identity, $\partial_\mu A_\mu = 2  \rho P$ 
($P$~=~$\bar q \gamma_5 q$). The non-perturbatively evaluated renormalization constants  are:
$Z_{A}^{(0)}~=~0.81(1)$~\cite{LANL,alpha-APE}  and $\tilde b_A = 1.19(6)$~\cite{LANL} at $\beta=6.2$.
We have checked that the inverse lattice spacing obtained from the pion decay constant is 
consistent with the one we reported in our previous paper. Since the temporal extension of 
the lattice used in our previous paper is much larger than the present one we will use, whenever needed,   
$a^{-1}(f_\pi) = 2.71(12)~\gev$ from ref.~\cite{gdd}.

\begin{figure}[htbp]
\vspace*{-.1cm}
\begin{center}
\begin{tabular}{@{\hspace{-0.7cm}}c}
\epsfxsize11.0cm\epsffile{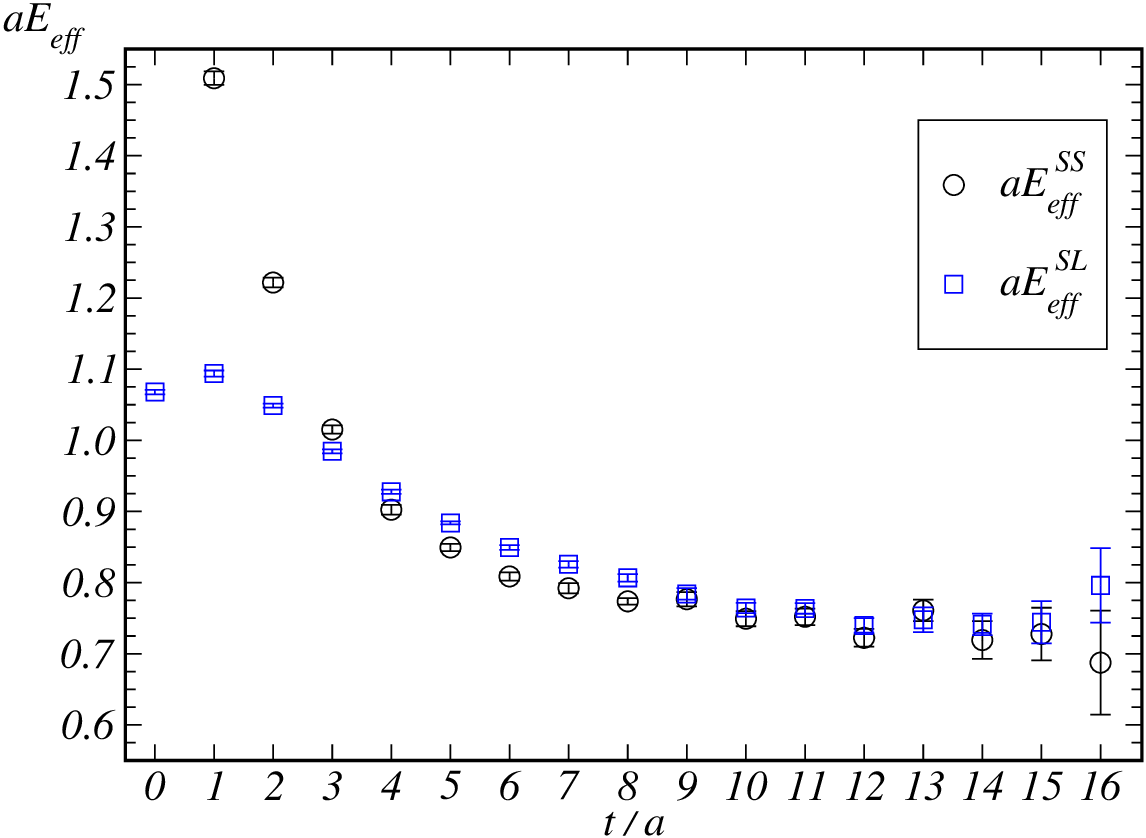}   \\
\end{tabular}
\caption{\label{meff}{\small\sl Signals for the effective binding energies obtained 
by using eq.~(\ref{Zl}) in case of Smeared-Smeared (``$SS$") and Smeared-Local (``$SL$") 
sources. In both cases the light quark mass corresponds to $\kappa_q=0.1344$, whereas for 
the static quark we use the ``fat" links as discussed in the text (see eq.~(\ref{Wilstat})]).}}
\end{center}
\end{figure}

{\it Static heavy--light correlators:} The binding energy $E_H$ and the constant ${\cal Z}$ 
are obtained from the two-point correlation function as indicated in  eq.~(\ref{CalZ1}). 
To identify the plateau in time where the ground state is properly isolated, we study the 
effective binding energy,
\bea \label{Zl}
E^{SS}_{\rm eff}(t) &=& \ln \left( \frac{C_2^{SS}(t)}{C_2^{SS}(t+1)} \right)\; .
\eea
The superscript ``$SS$" refers to the situation in which both interpolating fields are smeared. 
Note also that the projection 
$\frac{1+\gamma_0}{2}$ of the static quark onto the positive energy states reduces the
contamination of the signal coming from the opposite side of the periodic lattice. 
Therefore, the time reversal symmetry over the time dimension of the lattice
is not preserved in the heavy quark limit and the two-point
functions are no longer symmetric with respect to $t=T/2$.
 We illustrate in fig.~\ref{meff} the signals for the effective binding energies 
of the $C_2^{SS}$ and $C_2^{SL}$ correlators.
By fitting to the form~(\ref{CalZ1}) in the time interval in which the plateau is observed, we
extract $E_H$ and ${\cal Z}$. 
We have also considered a fit with two  exponentials, but did not observe any noticeable 
effect. The double exponential fit becomes necessary only for when performing simulations without 
the ``fattening" indicated in eq.~(\ref{Wilstat}) because in that case 
the plateaus of $E_{\rm eff}(t)$ are much shorter.

We first calculate the $B$-meson decay constant, 
$f_{B_s}^{\rm static}$. Rather than  providing an accurate value for 
this quantity (a precision determination of $f_{B_s}^{\rm static}$, has already been 
made in ref.~\cite{DellaMorte:2003mn}), our intention is only to test the validity of our computation. 
${\cal Z}^{L}$ is not obtained directly from the local--local correlators because the corresponding 
correlators are much less stable and statistically noisier than the smeared--local ones. For this 
reason ${\cal
Z}^{L}$ is extracted from  
\bea
\label{RZ} R_{\cal Z}(t) &=& \ {\cal Z}^S \times \left. \frac{C_2^{SL}(t)}{C_2^{SS}(t)} \right|_{\,t \gg 1} \, \to \,
{\cal Z}^L \; .
\eea
In table~\ref{tab_Hl}, we present the values of $E_H^{SS}$, $E_H^{SL}$, ${\cal Z}^{S}$, and ${\cal
Z}^{L}$ for a static heavy quark and  light quarks with masses corresponding to the $\kappa_{q_i}$'s 
specified in eq.~(\ref{eq0}).
\begin{table}[ht]
\begin{center}
\hspace*{-1cm}
\begin{tabular}{|c|c c c c|}
\hline
{\phantom{\Huge{l}}}\raisebox{-.2cm}{\phantom{\Huge{j}}}
\hspace*{-7mm}&  $aE_H^{SS}$ & $aE_H^{SL}$ &  ${\cal Z}^{S}$ &  ${\cal Z}^{L}$\\  \hline
{\phantom{\Huge{l}}}\raisebox{-.2cm}{\phantom{\Huge{j}}}
\hspace*{-7mm} static--${q_1}$
& $0.760(4)$ & $0.758(4)$ & $64.8 \pm 2.3$ & $0.126(3)$ \\
{\phantom{\Huge{l}}}\raisebox{-.2cm}{\phantom{\Huge{j}}}
\hspace*{-7mm} static--${q_2}$
& $0.748(5)$ & $0.745(5)$ & $60.8 \pm 2.4$ & $0.121(3)$ \\
{\phantom{\Huge{l}}}\raisebox{-.2cm}{\phantom{\Huge{j}}}
\hspace*{-7mm} static--${q_3}$
& $0.740(5)$ & $0.737(6)$ & $57.8 \pm 2.5$ & $0.117(3)$  \\ \hline
\end{tabular}
\caption{\label{tab_Hl}
\small{\sl The values of the binding energies obtained by using the Smeared-Smeared 
and Smeared-Local two-point correlation functions. The corresponding constants 
${\cal Z}^{S}$ and ${\cal Z}^{L}$ are also given (see eq.~(\ref{CalZ1})). The time intervals
chosen for the fits are: $t/a\in
[8,13]$ for $E_H^{SS}$, ${\cal Z}^{S}$, and $t/a\in
[9,13]$ for $E_H^{SL}$, ${\cal Z}^{L}$.}}
\end{center}
\vspace*{-.3cm}
\end{table}
The decay constant $f_{B_s}^{\rm static}$ is obtained from ${\cal Z}^{L}$
using  $m_{B_s}=5.37$~GeV, i.e. the physical $B_s$-meson mass, 
\bea \label{fB}
f_{B_s}^{\,static} &=& Z_A^{\,stat} \, {\cal Z}^{L} \, \sqrt{\frac{2}{m_{B_s}}} \, a^{-3/2}\, , 
\eea
and $Z_A^{\,stat} = 0.77(3)$,  the static heavy--light 
axial current renormalization constant computed non-perturbatively in ref.~\cite{Heitger:2003xg}. 
Notice that this value of $Z_A^{\,stat}$ is obtained by using the Eichten--Hill static heavy 
quark action, i.e. without including the effects of the fattening~(\ref{Wilstat}). 
In ref.~\cite{DellaMorte:2003mn}, however, it has been argued, on the basis of the observed behaviour of the step
scaling function, that the effect on $Z_A^{\,stat}$  of the modification of the static quark action is only very small. Therefore the  value we are using is certainly sufficient to achieve our present goal which, we recall, is to check the compatibility of our results with previous ones.

\begin{figure}[htbp]
\vspace*{-.1cm}
\begin{center}
\begin{tabular}{@{\hspace{-0.7cm}}c}
\epsfysize5.6cm\epsffile{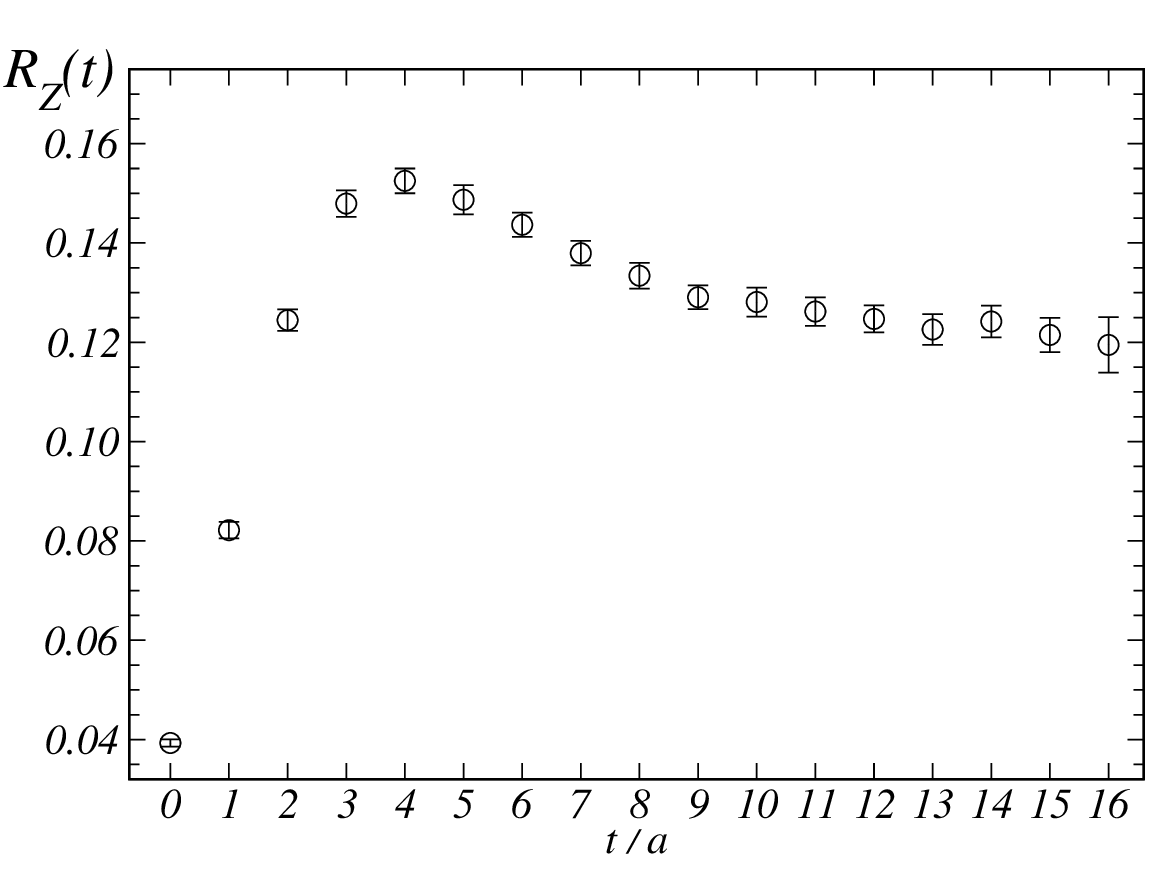}  \hspace*{0.3cm}  \epsfysize5.6cm\epsffile{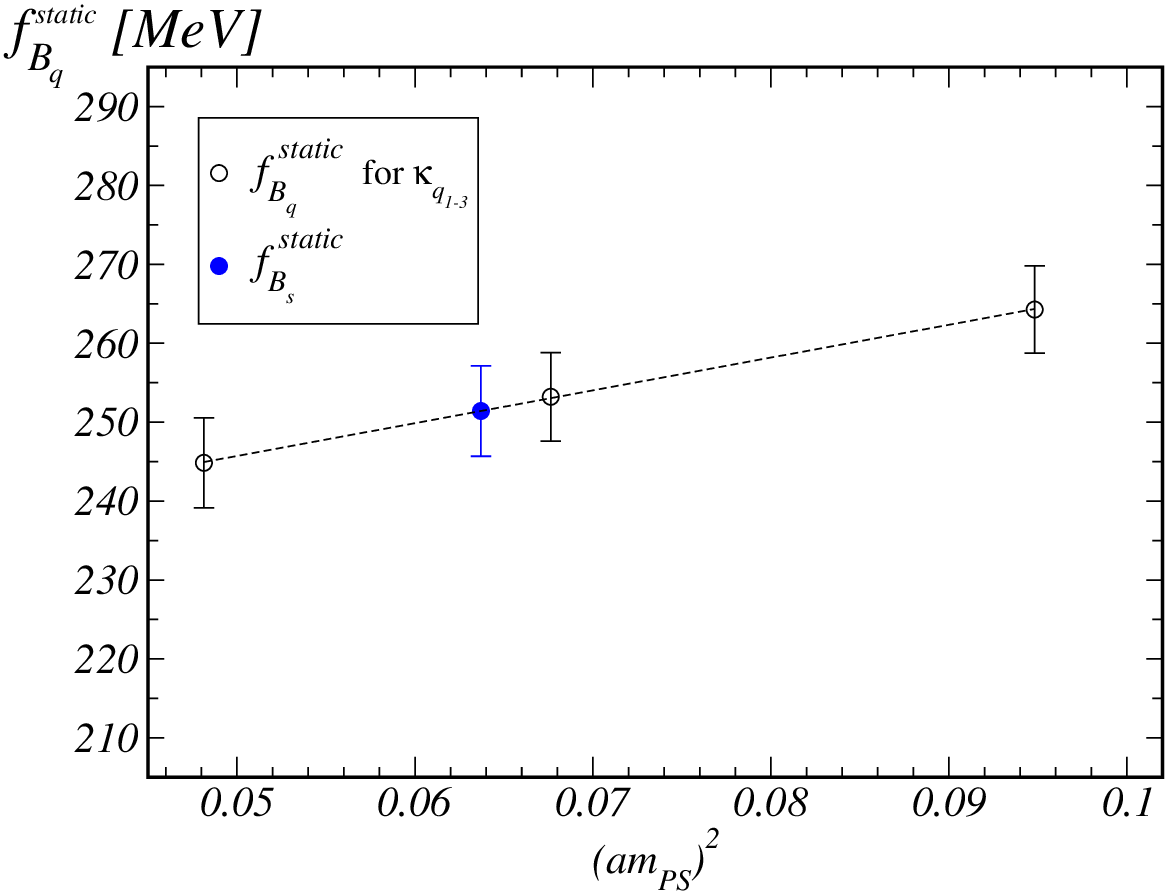}\\
\end{tabular}
\caption{\label{fBf}{\small \sl
 On the left, we illustrate the signal for the ratio $R_z(t)$ in the case 
$\kappa = 0.1344$. The fit of the plateau 
($t/a \in [9;13]$) to a constant gives ${\cal Z}^{L}$, which is then used to compute 
$f_{B_{q}}^{\rm static}$ (see eq.~(\ref{fB})). The corresponding 
decay constants are shown in the right plot (empty circles). The result of the 
linear interpolation to the strange light quark, i.e. $f_{B_{s}}^{\rm static}$, 
is marked by the filled circle. Notice that on the right plot only the statistical 
errors due to ${\cal Z}_L$ are displayed.}}
\end{center}
\end{figure}

In fig.~\ref{fBf}, we provide an illustration of the signal (plateau) of the ratio 
$R_z(t)$, and plot the corresponding $f_{B_q}^{\rm static}$, as obtained by using eq.~(\ref{fB}).
From the linear interpolation to the strange quark mass (corresponding to $m_{ss}^2=2 m_K^2 - m_\pi^2$), 
we obtain $f_{B_s}$, which we quote as
\bea
\label{fBsnum}
f_{B_s}^{\rm static} =\frac{ Z_A^{\,stat}}{ 0.77} \ (\ 251 \pm 20~\mev \,).
\eea
This value agrees quite well with the one,  $f_{B_s}^{\rm static} = 225 \pm 10~\mev$, recently obtained also 
at  $\beta =6.2$, in ref.~\cite{DellaMorte:2003mn}.

\subsection{Study of the three-point functions}

\hspace*{\parindent}We observe that, for $t_x$ not too close to 0 or $t_y$, the  three-point  function,  $C_{3  \, i i}^{SS}(0,t_x,t_y)$, introduced in
eq.~(\ref{C3}), does not depend on $t_x$. The reason for this is that, in the static limit, the ground states of the vector and pseudo-scalar
mesons are degenerate. Therefore, by writing 
$C_{3  \, i i}^{SS}(0,t_x,t_y)$ as
\bea
\sum_k\sum_l \left\langle 0\left|P(t_y)\right|H_k  \right\rangle\ \left\langle
H_k  \left|A_i(t_x)\right|H_l^\ast\right\rangle \ \left\langle H_l^\ast  \left|{V_i}^\dagger(0)\right|0\right\rangle\,,
\eea
where  $k$  and $l$ label the excitations, one sees that  in the  {\it diagonal} 
contribution ($k=l$) the exponential time dependence writes
\bea \label{Ediag}
e^{-(E_{H_l^\ast} - E_{H_k}) t_x -  E_{H_k} t_y} = e^{-E_{H_k} t_y} \;  ,
\eea
which is independent of $t_x$ ($t_y$ is kept fixed). The same is true for the ratio
(\ref{ratio}) in which the denominator is composed of two correlators 
$C_2$ which are degenerate in this case so that the exponential $t_x$
dependence  explicitly  cancels out,  and one ends up with
\bea   \label{ratio2}   R^{SS}(t_x)   =  \frac{  C_{3  \,
ii}^{SS}(t_x)  }{{{\cal Z}_S}^2 e^{-E_H^{SS} t_y}} \; .
\eea
 In other words, the plateau
for  the  ratio $R^{SS}(t_x)$ is equivalent to a plateau of 
$C_{3  \,  i  i}^{SS}(t_x)$.  A remarkable feature  is that $C_3$ already
develops   a   plateau  at  very a small  time  $t_x$.  The  physical
interpretation  could be that in the static limit, the {\it non-diagonal} matrix elements 
of  the axial current  simply vanish,
\bea \label{ndiag} \left\langle
H_k  \left|A_i(t_x)\right|H_l^\ast\right\rangle_{k\ne l} \sim 0 \ ,
\eea
 which implies that the  $t_x$-dependence in the l.h.s. of eq.~(\ref{Ediag}) vanishes as well when $k\,\neq\, l$.
Note that this is the case in a quark
model  framework where the axial current reduces to a  light  quark  spin  operator. 
Since the spatial wave functions  of  the ground  state and the excited states are orthogonal, the
spin  operator does verify  eq.~(\ref{ndiag})~\footnote{ This statement is verified by our 
results using the static heavy quark. It is further 
corroborated by the matrix elements obtained by using the propagating heavy quark 
with a mass around the physical charm quark: we see that, as the heavy
quark mass is increased, the three-point function $C_{3 \, i i}^{SS}(0,t_x,t_y)$ begins to
develop a plateau for smaller and smaller  values of $t_x$.}.

\begin{figure}[htbp]
\vspace*{-.1cm}
\begin{center}
\begin{tabular}{@{\hspace{-0.7cm}}c}
\epsfxsize11.0cm\epsffile{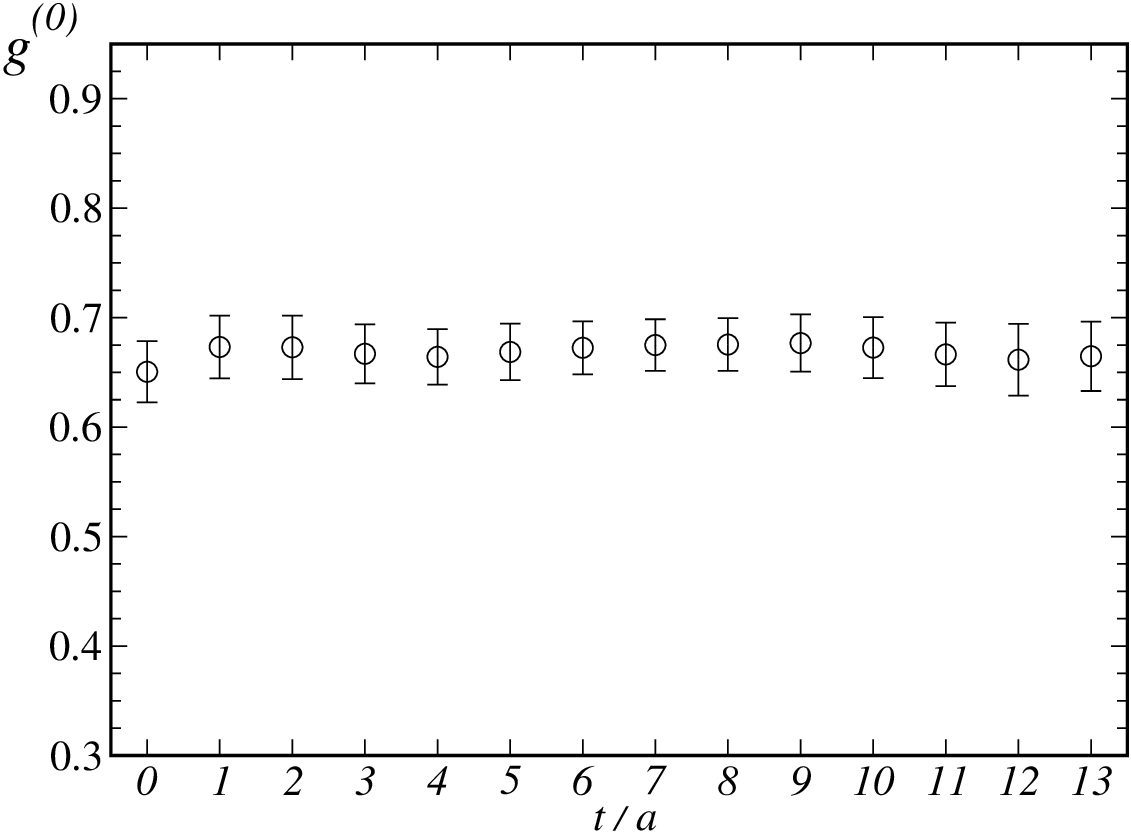}\\
\end{tabular}
\caption{\label{g0}{\small \sl Signal for the ratio $R^{SS}(t_x)$ (see eq.~(\ref{ratio2})). Illustration is
provided for $\kappa_{q_1} = 0.1344$. From the fit in $t/a \in [3,10]$, we obtain the value of
$\ghat_{\infty}^{(0)}$.}}
\end{center}
\end{figure}

\begin{figure}[htbp]
\vspace*{-.1cm}
\begin{center}
\begin{tabular}{@{\hspace{-0.7cm}}c}
\epsfxsize11.0cm\epsffile{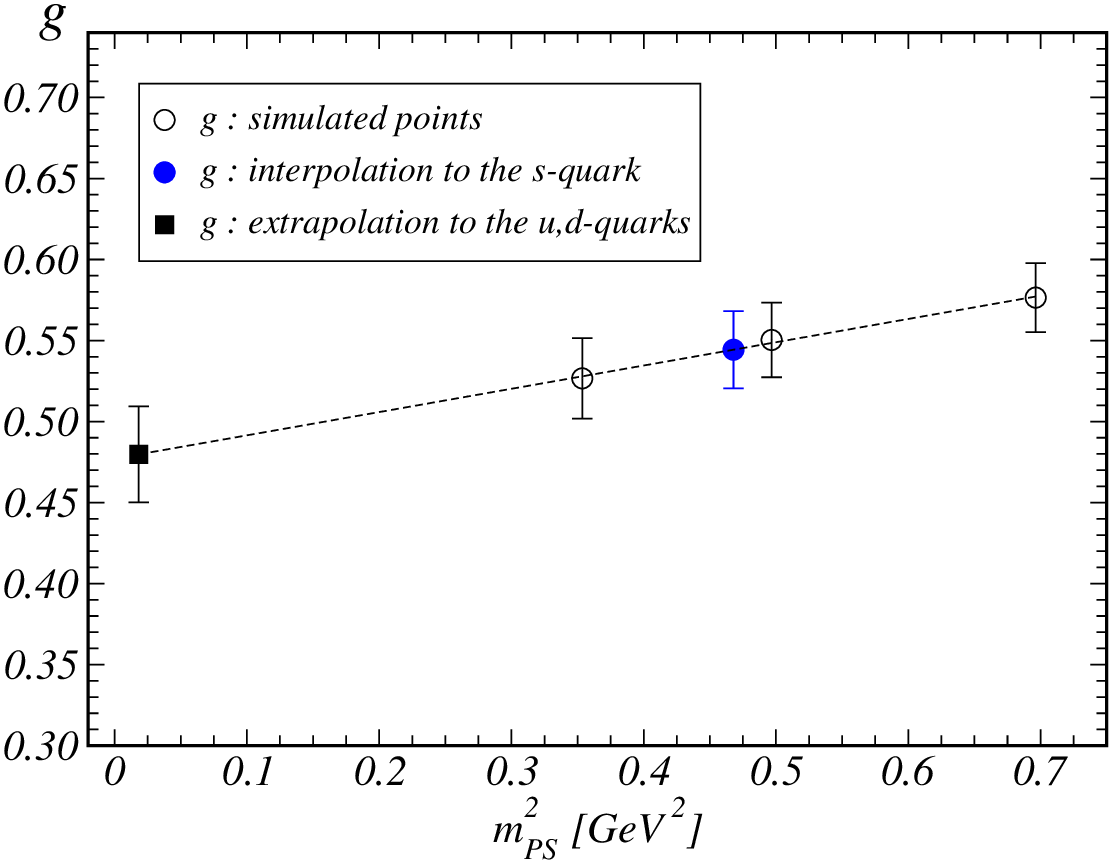}\\
\end{tabular}
\caption{\label{gchi}{\small \sl Linear extrapolation of $\ghat_{\infty}$ to the chiral limit. The empty circles
correspond to the simulated points whereas the filled symbols are the result of extra(inter)polation.}}
\end{center}
\end{figure}

As shown in eq.~(\ref{ratio}), the value of the bare
coupling $\ghat_{\infty}^{\,(0)}$, is obtained from the fit of $R^{SS}(t_x)$ to a constant. We  
illustrate the signal for $R^{SS}(t_x)$ in fig.~\ref{g0}. The corresponding results are listed in  
 table~\ref{tab_3p}, where we also give the final results
for  $\ghat_{\infty} = Z_A^I \, \ghat_{\infty}^{(0)}$, for our set of 
light quarks~(\ref{eq0}). We also include the values of $\ghat_\infty$ when the light quark 
is either the strange or the $u/d$ quark.
\begin{table}[ht]
\begin{center}
\hspace*{-1cm}
\begin{tabular}{|l|c c|}
\hline
{\phantom{\Huge{l}}}\raisebox{-.2cm}{\phantom{\Huge{j}}}
\hspace*{-7mm}&  $\ghat_{\infty}^{(0)}$ &  $\ghat_{\infty}$\\  \hline
{\phantom{\Huge{l}}}\raisebox{-.2cm}{\phantom{\Huge{j}}}
\hspace*{-7mm} static--${q_1}$
& $0.67(2)$ & $0.58(2)$  \\
{\phantom{\Huge{l}}}\raisebox{-.2cm}{\phantom{\Huge{j}}}
\hspace*{-7mm} static--${q_2}$
& $0.65(3)$ & $0.55(2)$  \\
{\phantom{\Huge{l}}}\raisebox{-.2cm}{\phantom{\Huge{j}}}
\hspace*{-7mm} static--${q_3}$
& $0.63(3)$ & $0.53(2)$  \\ \hline
{\phantom{\Huge{l}}}\raisebox{-.2cm}{\phantom{\Huge{j}}}
\hspace*{-7mm} static--$s$
& $0.64(3)$ & $0.54(3)$  \\ \hline
{\phantom{\Huge{l}}}\raisebox{-.2cm}{\phantom{\Huge{j}}}
\hspace*{-7mm} static--$u/d$
& $0.59(3)$ & $0.48(3)$  \\ \hline
\end{tabular}
\caption{\label{tab_3p}
\small{\sl Bare and renormalized values of $\ghat_\infty$. The time interval
chosen to fit for a plateau in the ratio $R^{SS}(t)$ is $t/a\in [3,10]$.}}
\end{center}
\vspace*{-.3cm}
\end{table}
From the linear extrapolation to the chiral limit\footnote{We defer to subsection~(\ref{syst43}) the discussion of the validity of this procedure and of the uncertainties it induces.},  shown in fig.~\ref{gchi}, we 
finally obtain, 
\bea
\label{ginfr}
\ghat_\infty = 0.48(3)\, .
\eea
The results presented in this section have been obtained after having fixed $t_y = 13a < T/2$.
This choice was made after we performed several simulations with $t_y/a \in [8,13]$ and  
studied $R^{SS}(t_x){\cal Z}_S^2$ for each $t_y$. We have checked that the result for $R^{SS}(t_x){\cal Z}_S^2$ 
does not change as soon as  $t_y$ gets larger than  $10a$. To separate the sources as far as possible we have chosen to fix $t_y$ at $13\,a$.

\section{Smearing, Fattening and Systematic uncertainties\label{syst}}

\hspace*{\parindent}As we already mentioned, to improve the statistical quality of the correlation functions computed with the 
static heavy quark action, we used the ``fattening" procedure~\cite{DellaMorte:2003mn}. In addition, we
use the smearing of the heavy--light interpolating fields~\cite{Boyle:1999gx}, which helps isolating the 
lowest bound state at lower time separations. In this section we explain how the parameters of the  smearing  
procedure were chosen~(\ref{Smearing})  and show the benefits of using the ``fat link" static 
heavy action instead of the standard Eichten--Hill one. After that discussion, 
we enumerate the sources of systematic uncertainties and estimate the related errors.

\subsection{\label{sme}Smearing}

\begin{figure}[htbp]
\vspace*{-.1cm}
\begin{center}
\begin{tabular}{@{\hspace{-0.7cm}}c}
\epsfxsize11.0cm\epsffile{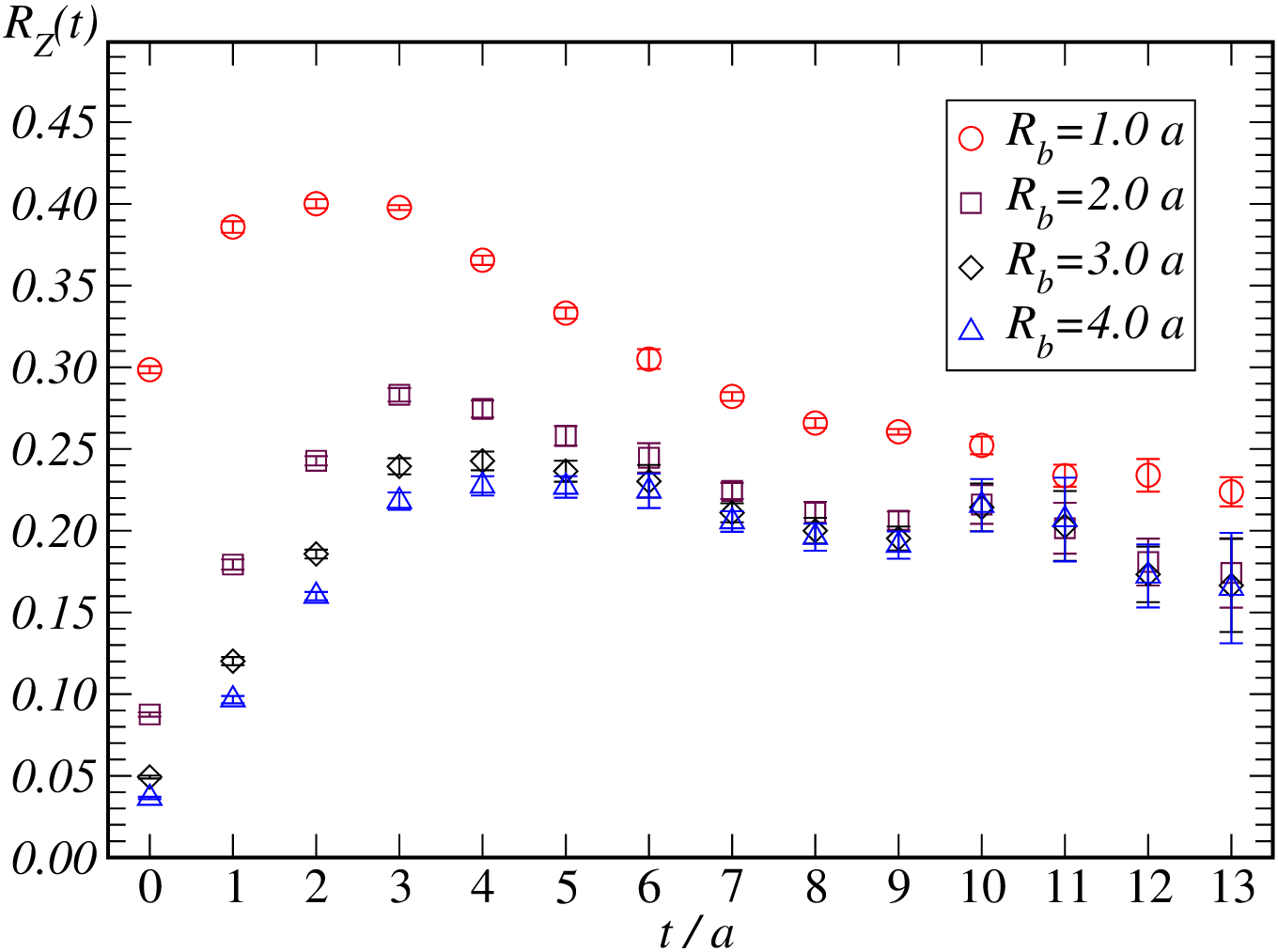}\\
\end{tabular}
\caption{\label{Rbf}{\small \sl Effect on the ratio $R_{\cal Z}(t)$ of the variation of $R_b$. For
$R_b > 2.0\,a$ the value of $R_{\cal Z}$ remains almost unchanged in the region t$\ge7\,a$. For this
illustration we used the Eichten--Hill static heavy quark action, while the light Wilson quark 
$\kappa$  is 0.1344.}}
\end{center}
\end{figure}

\hspace*{\parindent}The smearing procedure involves four parameters: $R_{\rm max}$, $R_b$, $N_F$ and $C$ 
(see eqs.~(\ref{Smearing}-\ref{Fuzzing})). The value of
$R_{max}$ represents half  the size of the hypercube where the smearing wave function lives and it 
is set  to the maximal value allowed by the memory limitations. In our case, $R_{max} = 5\,a$. 
The parameter $R_b$, which appears in the wave function $\phi({ r}) = e^{-r/R_b}$, 
is not fixed 
by any physical requirement other than improving the shape of the plateau for
$R_{\cal Z}$. The smeared--local correlators are much more sensitive 
to the variation of $R_b$ than the smeared--smeared ones. 
In fig.\ref{Rbf} we show the shape of $R_{\cal Z}(t)$ for $R_b \in [1.0\,a\, ,\, 4.0\,a]$.
We see that if $R_b\geq 3.0\,a$, the value of $R_{\cal Z}(t)$ on the plateau, 
and thus also the corresponding $f_{B_q}^{\rm static}$, remains unchanged. 
The reason is that, as the wave function
$\phi(r)$ lives in a typical volume of radius $R_{max} = 5\,a$, the shape of the
normalized wave function term [$(r+1/2)^2 \phi(r)$] in eq.~(\ref{Smearing}) does 
not vary much when changing $R_b$ in $3a-5a$. Of course, all the physical quantities
are built in such a way that, at the end, the effect of the normalization of
$\phi(r)$ (i.e. the unwanted effect of the smearing) cancels. On the basis of these observations, 
we fix $R_b=3a. $\footnote{
As a side remark, we
note that the physical Compton wavelength of the $D$-mesons $\lambda_D$  is
given by the mass splitting between the $D$ and the $D_1$ mesons masses. Its value,
$\lambda_D = 2.2~{\rm fm}$, is somewhat similar to 
$R_b=3.0\,a \sim 2~{\rm fm}$, that we use in our simulation.}

The remaining two parameters, $C$ and $N_{F}$, enter in the fuzzing procedure~(\ref{Fuzzing}). 
As far as $C$ is concerned, it turns out that its value does not make any significant 
effect on our result. We varied its value in 
the range $C\in [0.8\, , \,4.0]$, and the shape of $R_{\cal Z}(t)$ does not change at all 
 (of course, within our present statistical accuracy). On the other
hand, the number of fuzzing iterations, $N_{F}$, is chosen to be $N_{F} = 5$. 
The effect of each iteration is to fuzz the spatial links by the gauge fields in the 
immediate neighborhood. By several iterations we can restore the effect of the wave 
function inside the volume of radius $R_b$, where $\phi(r)$ is supposed to live. As $N_{F}$
 iterations cover a volume of typical radius $(N_{F}+1)^{1/2}\,a$, we should 
require $N_{F} \sim 8$ to fill a volume of radius  $R_b = 3.0\,a$. It turns out, however, that 
for $N_F \geq 6$, 
the signal for $R_{\cal Z}$ begins to increase sharply: a large number of fuzzing iterations 
destroy the quantum fluctuations as it was observed in the cooling procedures used to 
study instantons~\cite{teper}. We checked that $R_{\cal Z}$, and thus $f_B^{\rm static}$,  
remain stable for  $N_F \in [3,5]$, and for the final simulation we chose $N_F=5$.

\subsection{Effect of  ``fattening" the static heavy quark propagator}

\begin{figure}[htbp]
\vspace*{-.1cm}
\begin{center}
\begin{tabular}{@{\hspace{-0.7cm}}c}
\epsfxsize7.9cm\epsffile{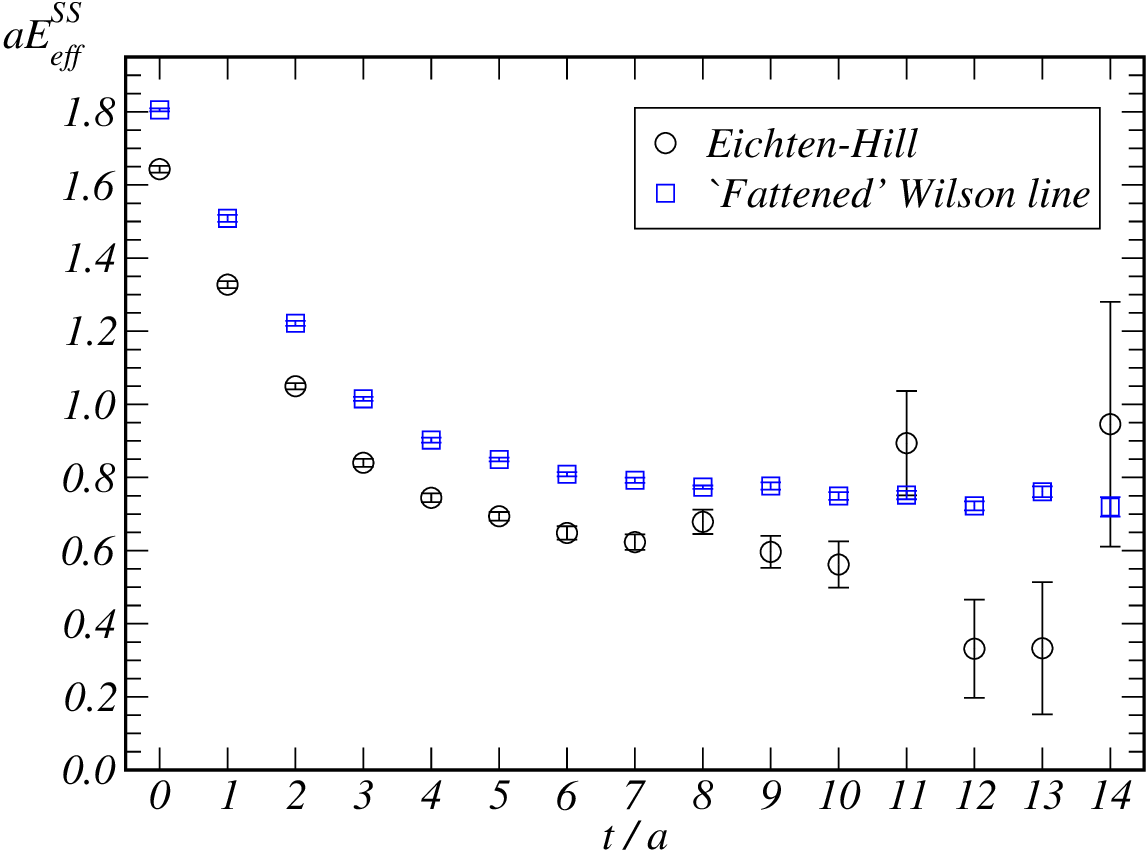} \hspace*{0.3cm} \epsfxsize7.9cm\epsffile{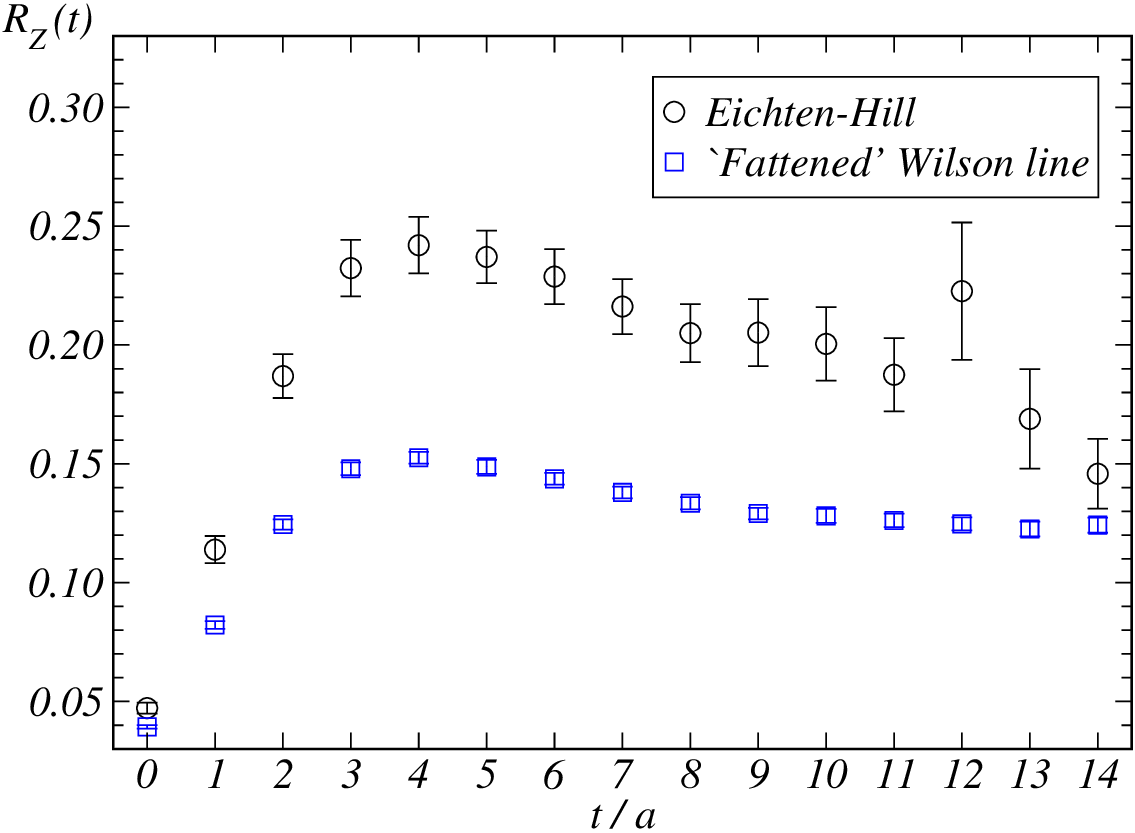}\\
\end{tabular}
\caption{\label{fat}{\small \sl  Comparison of the signals for the effective binding energies 
$E_{eff}^{SS}$ and
${\cal Z}^{L}$ when using either the usual Eichten--Hill action or its ``fattened" version (see 
eq.~(\ref{Wilstat})).} }
\end{center}
\end{figure}

\hspace*{\parindent}In this subsection we show that replacing the link variables in the Eichten--Hill action 
by the ``fat" ones (see eq.~(\ref{Wilstat})), does make the plateau 
for the effective binding energies much larger, including  a visible reduction of the statistical noise. 
In fact, after replacing $U_t \to U_t^{\rm fat}$ in the Wilson line, the plateau gets extended to 
$t\sim 12\,a-13\,a$, which is actually the upper limit ($T/2=14\,a$). From the comparison of the effective 
energy, eq.~(\ref{Zl}), obtained by using the ``fat" Wilson line and of the standard one (i.e. without fattening),
we can clearly see the improvement in the signal. This is illustrated in fig.~\ref{fat}, where in 
the case of ``fat" Wilson line, the statistics is half what it was in the other case. This improvement is still more striking in the case where one uses the so-called Hyp-fattening (cf. refs.~\cite{Ahasen1,Ahasen2}) instead of the standard one (see ref.~\cite{Benoit}).
We see that not only the plateau appears much clearer, but the signal remains longer in time. 
We also see that the two values of the effective binding energies  are different, as 
expected, since the heavy quark lattice action was modified.
In table~\ref{Comp_fat}, we compare the results obtained in both cases, i.e. 
with and without using the fattening procedure. 
\begin{table}[ht]
\begin{center}
\hspace*{-1cm}
\begin{tabular}{|c|c c|}
\hline
{\phantom{\Huge{l}}}\raisebox{-.2cm}{\phantom{\Huge{j}}}
\hspace*{-7mm}&Eichten--Hill& `Fattened' \\  \hline
{\phantom{\Huge{l}}}\raisebox{-.2cm}{\phantom{\Huge{j}}}
\hspace*{-7mm} $f_{B_s}^{\rm static}~[\mev]$
& $410(56)$ & $251(20)$  \\
{\phantom{\Huge{l}}}\raisebox{-.2cm}{\phantom{\Huge{j}}}
\hspace*{-7mm} $\ghat_{\infty}$
& $0.58(8)$ & $0.50(3)$  \\ \hline
\end{tabular}
\caption{\label{Comp_fat} \small{\sl Values of 
$f_{B_s}^{\rm static}$ and $\ghat_{\infty}$ as computed from the usual
Eichten--Hill static quark action and the new procedure with ``fat" links. 
Parameters of both simulations are kept identical. In particular, to determine
$\ghat_\infty$ we have taken in this case $t_y = 10\,a$.}}
\end{center}
\vspace*{-.3cm}
\end{table}

Before closing this subsection, we note also that, as far as the stability of the signal is concerned, 
the effect of replacing the standard Eichten--Hill action 
by the fat-link one is much more rewarding in the case of the heavy--light decay constant than for our 
$\ghat_\infty$. This is so because the operator whose 
matrix element is related to $f_{B}^{\rm static}$ includes a static quark, whereas the one related to 
$\ghat_\infty$  does not. The fat link action however helps improving the statistical quality 
of the signal for $\ghat_\infty$. We note that, besides this improvement, the use of fattened links leads to a dramatic decrease of the value of $f_{B_s}^{\rm static}$ with respect to what had been found with the Eichten-Hill action. The fact that the plain EH formalism faces problems when trying to study $f_{B}$ has been known for some time. In refs.~\cite{Lepage,Labrenz} it was shown that the
smearing led to systematically lower values\footnote{We thank the anonymous referee for drawing those references to our attention.}.  We show here that a similar effect results from the use of fattened links. In ref.~\cite{Benoit} we show that the Hyp-fattening confirms fully the present results.

\subsection{Systematic errors\label{syst43}}

\begin{itemize}

\item {\it Smearing}:

We combine in quadrature the effects of the
variation of all the parameters of the smearing procedure 
on the resulting value for $\ghat_\infty$ (see discussion in sec.~\ref{sme}). 
We estimate that systematics to be $\pm 15\%
$.

\item {\it Discretization errors}:

In our study we implemented the full  ${\cal O}(a)$ improvement of the Wilson QCD action and the axial current.
As we discussed in the text, the improvement of the bare axial current  does not influence the value of  $\ghat_\infty$.
As for the renormalization constant, we used the non-perturbatively determined  value, including the coefficient
$\widetilde b_A$, which ensures the elimination of the artifacts of ${\cal O}(a\rho)$.

Our main result is obtained from the simulation at $\beta = 6.2$. In order to study the ${\cal O}(a)$ effects,  
we also performed the simulations at $\beta =6.1$ and $\beta =6.0$,  keeping the physical volume approximately the same  
and  rescaling the smearing parameters. In fig.~\ref{g_60_62}, we show the chiral behavior of 
$\ghat_\infty$, as computed in all three simulations. 
\begin{figure}[htbp]
\vspace*{-.1cm}
\begin{center}
\begin{tabular}{@{\hspace{-0.7cm}}c}
\epsfxsize11.0cm\epsffile{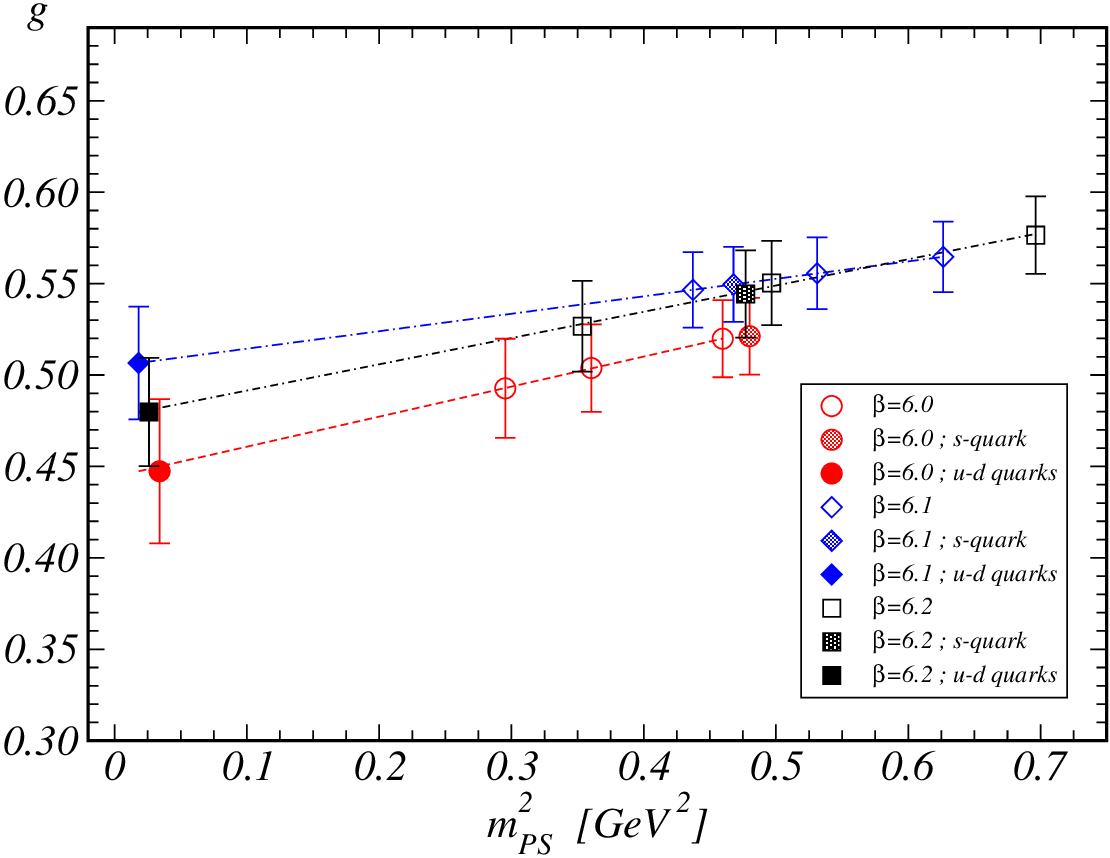}\\
\end{tabular}
\caption{\label{g_60_62}{\small\sl Comparison of the chiral extrapolations of $\ghat_\infty$ from three
simulations at $\beta = 6.0$, $6.1$ and $6.2$. The values extracted from these  simulations are compatible within one standard
deviation.}}
\end{center}
\end{figure}
We see that both the results accessed directly from our simulations and those obtained after 
the linear extrapolation agree within the error bars. To give the reader a more quantitative insight into this effect, 
 we collect in table~\ref{tab_60_62} the results for $\ghat_\infty$ and $f_{B_s}^{\rm static}$ from all 
three simulations.

\begin{table}[ht]
\begin{center}
\hspace*{-1cm}
\begin{tabular}{|c|c c c|}
\hline
{\phantom{\Huge{l}}}\raisebox{-.2cm}{\phantom{\Huge{j}}}
\hspace*{-7mm}$\beta$ & $ 6.0$ & $ 6.1$& $ 6.2$ \\  \hline
{\phantom{\Huge{l}}}\raisebox{-.2cm}{\phantom{\Huge{j}}}
\hspace*{-7mm} $f_{B_s}^{\rm static}~[\mev]$
& $250(35)$& $257(27)$  & $251(20)$  \\
{\phantom{\Huge{l}}}\raisebox{-.2cm}{\phantom{\Huge{j}}}
\hspace*{-7mm} $\ghat_{\infty}$
& $0.45(4)$ & $0.51(3)$ & $0.48(3)$ \\ \hline
\end{tabular}
\caption{\label{tab_60_62} {\small\sl Values of $f_{B_s}^{\rm static}$ and $\ghat_{\infty}$ from three simulations at $\beta =6.0$, $6.1$ and $6.2$.
Consistent results are found when comparing these  studies.}}
\end{center}
\vspace*{-.3cm}
\end{table}
Since in this work we do not deal with  propagating heavy quarks (which are usually the dominant source of discretization errors 
in the heavy--light meson observables), and since we do not observe any noticeable discretization error 
on $\ghat_\infty$ from our simulations at three different lattice spacings, we will not account 
for any extra discretization error in our overall systematic uncertainty\footnote{Notice also that  
$\ghat_\infty$ is dimensionless so that no discretization errors due to the conversion from 
the lattice to physical units arise.}.

\item {\it Finite volume}:

In this work we did not carry out a detailed study of the finite volume effects.  Instead,
we will rely on our previous study with the propagating heavy quark where the finite volume 
effects are estimated to be $\pm 6$\%~\cite{gdd}.
This uncertainty will be added in our overall systematic error estimate.

\item {\it Chiral extrapolations}:

Since we have only three values for the bare light-quark masses, we cannot make 
a detailed study of the effect of the chiral extrapolation like the one 
we did in ~\cite{gdd}. 
The values of  $\ghat_\infty$ that are directly accessed from our simulations 
follow a very smooth linear behavior when the value of the light quark mass is changed. 
The result~(\ref{ginfr}) is obtained from the linear extrapolation. To that we will 
add an error of $\pm 15\%
$, as estimated in our previous paper~\cite{gdd}. This error is considered symmetric for the reasons 
explained in detail in ref.~\cite{jsd}: the chiral logs in the full (unquenched) theory 
have a tendency to lower the value of the coupling $\ghat_\infty$, whereas those in 
the quenched theory do the opposite (increase the value $\ghat_\infty$).

\end{itemize}

After combining the above sources of systematic uncertainty in quadrature, we end up with $22\%$ 
of error to the value given in eq.~(\ref{ginfr}), i.e.
\bea
\ghat_\infty = 0.48 \pm 0.03\pm 0.11\;.
\eea

\section{Heavy mass interpolation to the $b$-quark\label{interp}}

\begin{figure}[htbp]
\vspace*{-.1cm}
\begin{center}
\begin{tabular}{@{\hspace{-0.7cm}}c}
\epsfxsize11.0cm\epsffile{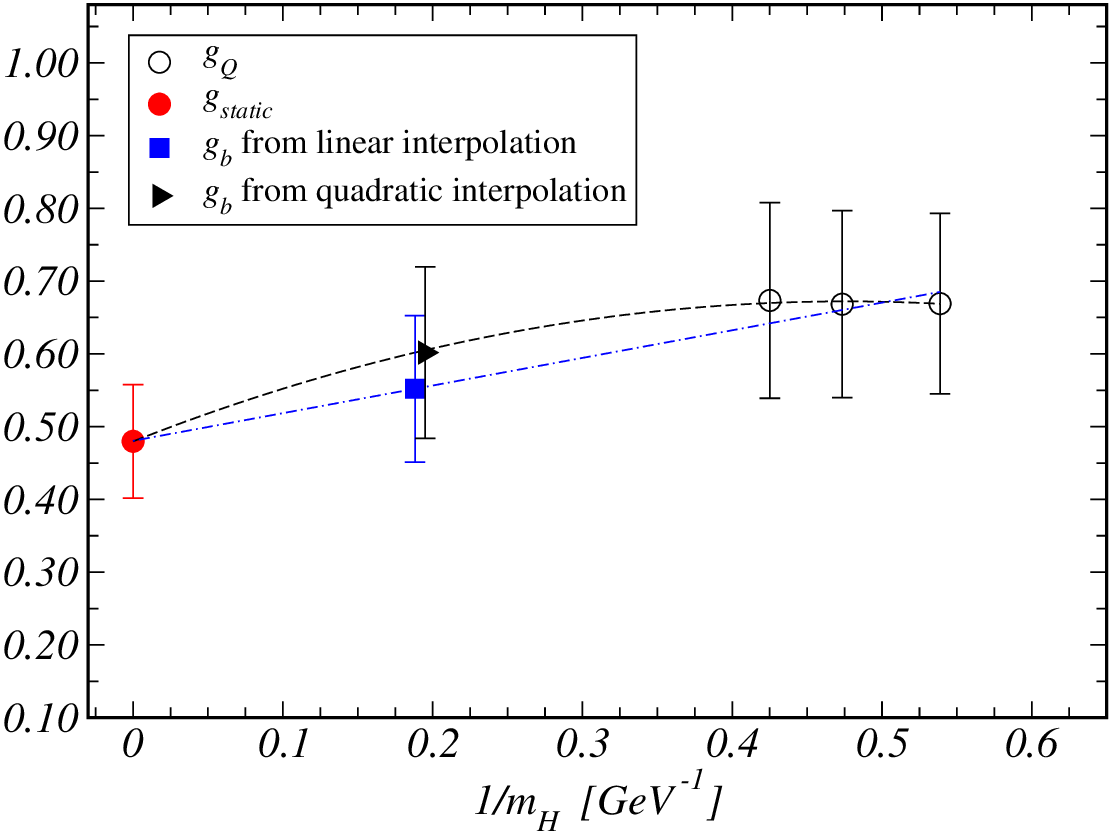}\\
\end{tabular}
\caption{\label{gbf}{\small \sl  Heavy mass interpolation of the coupling $\ghat$ to the $B$-meson sector, using
heavy masses around the charm (empty circles) and our new result at the static limit (filled circle). The linear
and quadratic fits defined in eqs.~(\ref{fits}) and (\ref{fitsq})  give compatible results for $\ghat_b$ (filled square and
triangle).}}
\end{center}
\end{figure}

\hspace*{\parindent}We can now combine our static heavy quark result, $\ghat_{\infty}$, with those obtained 
in our previous study in which propagating heavy quarks with masses around the physical 
charm quark were used~\cite{gdd}, to interpolate to the $b$-quark sector. 
For that purpose we use the spin-averaged mass of the heavy--light mesons, i.e.
\bea
 \overline{ m }_H = {\frac{3 m_V + m_P}{4}}\;.
\eea
Motivated by the heavy quark symmetry, we fit our results to the linear and quadratic forms: 
\bea \label{fits}
\ghat_{Q} &=& \ghat_\infty + \frac{a_1}{\overline{ m }_H} \;,
\eea
\bea\label{fitsq}
\ghat_{Q} &=& \ghat_\infty + \frac{b_1}{\overline{ m }_H} +
\frac{b_2}{\overline{ m }^2_H}\;.
\eea
These fits are shown in fig.~\ref{gbf}, from which we read off the values at 
$1/{\overline{ m }_B}=(5.314 \ \gev)^{-1}$, i.e. $\ghat_b$:
\bea
\ghat_b^{\ (lin)} = 0.55(4)(9), \quad \ghat_b^{\ (quad)} = 0.60(7)(9)\;.
\eea
As our final estimate we will take the average of the two, and add the difference in the overall 
systematics. Thus we have
\bea
\ghat_b = 0.58\pm 0.06 \pm 0.10\,,
\eea
which by means of eq.~(\ref{gg}) gives
\bea
\gbb = 47\pm 5 \pm 8\,.
\eea

\section{Summary and conclusions\label{sec4}}

\hspace*{\parindent}In this paper we made a study of $\ghat_\infty$, the coupling of the lowest-lying doublet 
of heavy--light mesons to a charged pion, in the infinitely heavy quark limit. 
The static heavy quark action used in this work is the  Eichten and Hill one with 
fat-links, which ensures a better statistical quality of the correlation functions 
computed on the lattice. Our result, in the quenched approximation, is
\bea
\ghat_\infty &=& 0.48 \pm 0.03\pm 0.11\,,
\eea
where the errors are statistical and systematic respectively. The most significant sources 
of systematic uncertainty are those related to the chiral extrapolation and to the smearing 
procedures used to suppress the contribution of the higher excited states in the relevant 
correlation functions.

That result is then combined with the ones reported in our previous work~\cite{gdd}, 
in which we used  propagating heavy quarks with  masses close to the one of the physical charm quark. 
Interpolation to the $b$-quark sector leads to
\bea
\ghat_b = 0.58\pm 0.06\pm 0.10\quad  {\rm and } \quad \gbb = 47\pm 5\pm 8\,.
\eea 

The quantities we have studied in this paper, $\hat{g}_\infty$ and $g_{B^* \,B\,\pi}$, are very interesting  because 
\begin {enumerate}
\item[1] they are of high phenomenological relevance,
\item[2] they escape direct measurement, 
\item[3] their theoretical determinations are very widespread.
\end{enumerate}
 
\noindent Still, as the attentive reader will have noticed, the errors remain larger large. Referring for instance to eq.~(28) above, the
 statistical error is  6\% and the systematic one 22\%. 

Concerning the statistical error a first factor is obviously the performance level
 of the 
available computing facilities.  Since this is getting higher and higher  rather fast and, at the same time, new computational techniques
  have been discovered which greatly improve the signal/noise ratio this source of uncertainty will probably be tamed 
very soon. Some of the systematic errors are no serious problems : most of the various renormalization and improvement constants we need have
now been non perturbatively  determined with an accuracy at the $\%$-level. Note also that some of the (less well-known) heavy-light constants (e.g. 
$Z_{B_s}^{static}$) simply drop out of our computation. Some other ones are more serious :
\begin{itemize} 
\item[] The chiral extrapolation (which we have
 very roughly estimated to contribute by 15\%) can be attacked on both the theoretical side (determination of the chiral logs)  and
the numerical one.
\item[] In this study we have used the static limit only as a limit point for our extrapolations from the  finite-mass region. It is
 possible to go beyond this approach  and to determine higher order $\frac{1}{M_Q}$
 coefficients, which would allow a kind of ``educated fit'' and to decresase the theoretical uncertainties. This is part of an ambitious
 program  which has been undertaken by the Alpha-group  3 years ago and is still under progress~\cite{alpha1,alpha2,alpha3,alpha4}.
\item[]  There remains the question of the unquenching. Its possibility while using the fattening procedures has been established (see 
ref.~\cite{AHasen}) at least in the staggered-fermions case  and probably with Wilson fermions too. This is one of the tasks we should undertake in the future.

\end{itemize}

\noindent These are just a few hints regarding what will have to  be done  now but it is impossible at  present 
to outline any really sensible quantitative perspective regarding the time evolution in the near future of the various uncertainty factors 
we have just mentionned. Still, pursuing in the direction of a ``pure QCD'' determination of $\widehat{g}$ is clearly of primary importance.

\vspace*{3cm}

\section*{Acknowledgement}

\hspace*{\parindent}We thank M.~Della Morte, J.~Reyes and A.~Shindler for discussions and comments. The simulation was
performed with the APE1000 located in the Centre de Ressources Informatiques (Paris-Sud, Orsay) and purchased
thanks to a funding from the Minist\`ere de l'Education Nationale and the CNRS. This work was supported in part
by the European Union Human Potential Program under contract HPRN-CT-2000-00/45, Hadrons/Lattice Phenomenology.

\end{document}